\documentclass[aps,prb,twocolumn]{revtex4}
\usepackage{epsfig,amsmath,amssymb}
\usepackage{color}

\newcommand{\be}{\begin{equation}}
\newcommand{\ee}{\end{equation}}
\newcommand{\ra}{\rangle}
\newcommand{\la}{\langle}
\newcommand{\bit}{\begin{itemize}}
\newcommand{\eit}{\end{itemize}}
\newcommand{\bea}{\begin{eqnarray}}
\newcommand{\eea}{\end{eqnarray}}

\newcommand{\Neel}{N\'{e}el}
\newcommand{\bfs}{{\bf s}}

\begin{document}

\title
{Ground-state phase diagram of the spin-1/2 square-lattice J1-J2 model with plaquette
structure}

\author
{O. G\"otze$^1$, S.E. Kr\"uger$^1$, F. Fleck$^1$, J. Schulenburg$^2$, and J. Richter$^1$}

\affiliation
{$^1$ Institut f\"ur Theoretische Physik, Universit\"at Magdeburg,
P.O. Box 4120, 39016 Magdeburg, Germany\\
$^{2}$Universit\"{a}tsrechenzentrum, Universit\"{a}t Magdeburg,
P.O. Box 4120, 39016 Magdeburg, Germany
}

\date{\today}

\begin{abstract}
Using the coupled cluster method for high orders of approximation
and Lanczos exact diagonalization we study the ground-state phase diagram
of a quantum spin-1/2 $J_1$-$J_2$ model on the square lattice with  plaquette structure.
We consider antiferromagnetic ($J_1>0$) as well as ferromagnetic ($J_1<0$) 
nearest-neighbor interactions together with frustrating antiferromagnetic 
next-nearest-neighbor interaction $J_2>0$. The strength of inter-plaquette
interaction $\lambda$ varies between $\lambda=1$ (that corresponds to the 
uniform $J_1$-$J_2$ model) and $\lambda=0$ (that corresponds to isolated
frustrated 4-spin plaquettes).
While on the classical level ($s \to \infty$) both versions of models (i.e., with ferro- and antiferromagnetic $J_1$)
exhibit the same ground-state behavior, the ground-state phase diagram
differs basically for the quantum case $s=1/2$.
For the antiferromagnetic case ($J_1 > 0$) \Neel\ antiferromagnetic
long-range order at small $J_2/J_1$ and $\lambda \gtrsim 0.47$ as well as
collinear striped
antiferromagnetic  long-range order at large $J_2/J_1$ and  $\lambda \gtrsim
0.30$ appear which  
correspond to
their classical counterparts.
Both semi-classical magnetic phases are separated by a nonmagnetic quantum paramagnetic
phase. The parameter region, where this nonmagnetic phase exists, increases with decreasing of
$\lambda$.
For the ferromagnetic case ($J_1 <  0$) we have the trivial ferromagnetic
ground state at small $J_2/|J_1|$. By  increasing of $J_2$ this classical phase gives way for 
a semi-classical plaquette phase, where the plaquette block spins of length
$s=2$  are antiferromagnetically long-range ordered.
Further increasing of $J_2$ then yields
collinear striped
antiferromagnetic  long-range order for $\lambda \gtrsim 0.38$, but a
nonmagnetic quantum paramagnetic
phase  $\lambda \lesssim 0.38$.
\end{abstract}


\maketitle

\section{Introduction}
\label{intro}
The spin-1/2 quantum Heisenberg antiferromagnet with nearest-neighbor (NN),
$J_1 > 0$,  
and next-nearest-neighbor (NNN) bonds, $J_2\geq 0$, on the square lattice has attracted 
much interest (see, e.g., Refs.~\onlinecite{chandra88,dagotto89,schulz,richter93,richter94,
zhito96,Trumper97,bishop98,singh99,sushkov01,capriotti01,Sir:2006,Schm:2006,darradi08,
sousa,bishop08,xxz,ortiz,singh2009,cirac2009,ED40,fprg,cirac2010,balents2011,verstrate2011}) 
as  a canonical model to study the 
interplay between frustration 
and quantum fluctuations. In particular, the quantum phase transitions inherent
in this model as well as the nature of its quantum paramagnetic
phase in the region $0.4\lesssim J_2/J_1 \lesssim 0.6$ is a matter of intensive debate.
The \Neel\ antiferromagnetic  (NAF) long-range order (LRO) at small $J_2/J_1$
and the  collinear striped  
antiferromagnetic  (CAF) LRO  at large $J_2/J_1$ correspond to
their classical counterparts, however, the sublattice magnetization is
reduced by quantum fluctuations.

Motivated by recent investigations on quasi-two-dimensional
frustrated magnetic materials with ferromagnetic (FM) NN
bonds
e.g.,
Pb$_2$VO(PO$_4$)$_2$,\cite{kaul04,jmmm07,carretta2009,enderle,nath}
 SrZnVO(PO$_4$)$_2$,\cite{rosner09,rosner09a,enderle,carretta2011} 
BaCdVO(PO$_4$)$_2$,\cite{nath2008,carretta2009,rosner09}
the FM $J_1$-$J_2$ model ($J_1<0, J_2\ge0$) has recently been
studied.\cite{shannon04,shannon06,sindz07,schmidt07,schmidt07_2,sousa,shannon09,
sindz09,haertel10,richter10,momoi2011,cabra2011} 
Although on the classical level the ground state (GS) phase diagrams of both variants of the $J_1$-$J_2$ model are
quite similar, the GS properties of the quantum  model with $J_1<0$
are basically different from those for $J_1>0$.     
Currently, the GS properties of the FM model 
around  $J_2/|J_1|=0.5$ are under controversial debate.
On the one hand in Refs.~\onlinecite{shannon06} and \onlinecite{momoi2011} arguments for a nematic
phase separating the classical FM and the semi-classical CAF
phases are given, on the other hand  
in Refs.~\onlinecite{richter10} and \onlinecite{cabra2011}        
an intermediate quantum paramagnetic GS phase 
is not found (or it exists in a very small
parameter region around $J_2\sim 0.4|J_1|$, only).
Hence, a direct first-order transition between the 
FM GS and the GS with CAF LRO at $J_2\sim
0.4|J_1|$ could take place.\cite{richter10,cabra2011}

Quantum phase transitions can also occur by competition between
antiferromagnetic (AFM) NN bonds 
of different strength, i.e., without frustration. One  example  
is the local singlet formation in dimerized Heisenberg models, see e.g.
Ref.~\onlinecite{hjs} and references therein.
As a certain extension to dimerized models the square-lattice Heisenberg model with a plaquette structure
has been considered,\cite{singh99,voigt02,ueda07,fabricio08,wenzel08,wenzel09,fledderjohann09}
where
local quadrumer singlet formation can destroy NAF LRO.

However, for the description of real materials a modification of the  $J_1$-$J_2$ model  
might be necessary. 
For example, for 
PbZnVO(PO$_4$)$_2$
a spatially anisotropy model has been
derived,\cite{rosner2010} whereas   
for 
(CuCl)LaNb$_2$O$_7$\cite{kageyama05} a $J_1$-$J_2$  model with additional plaquette structure
was proposed in Ref.~\onlinecite{ueda07}. This
plaquette model, however,  has been questioned recently by  Rosner and
coworkers.\cite{rosner_co}

\begin{figure}[ht]
\begin{center}
\epsfig{file=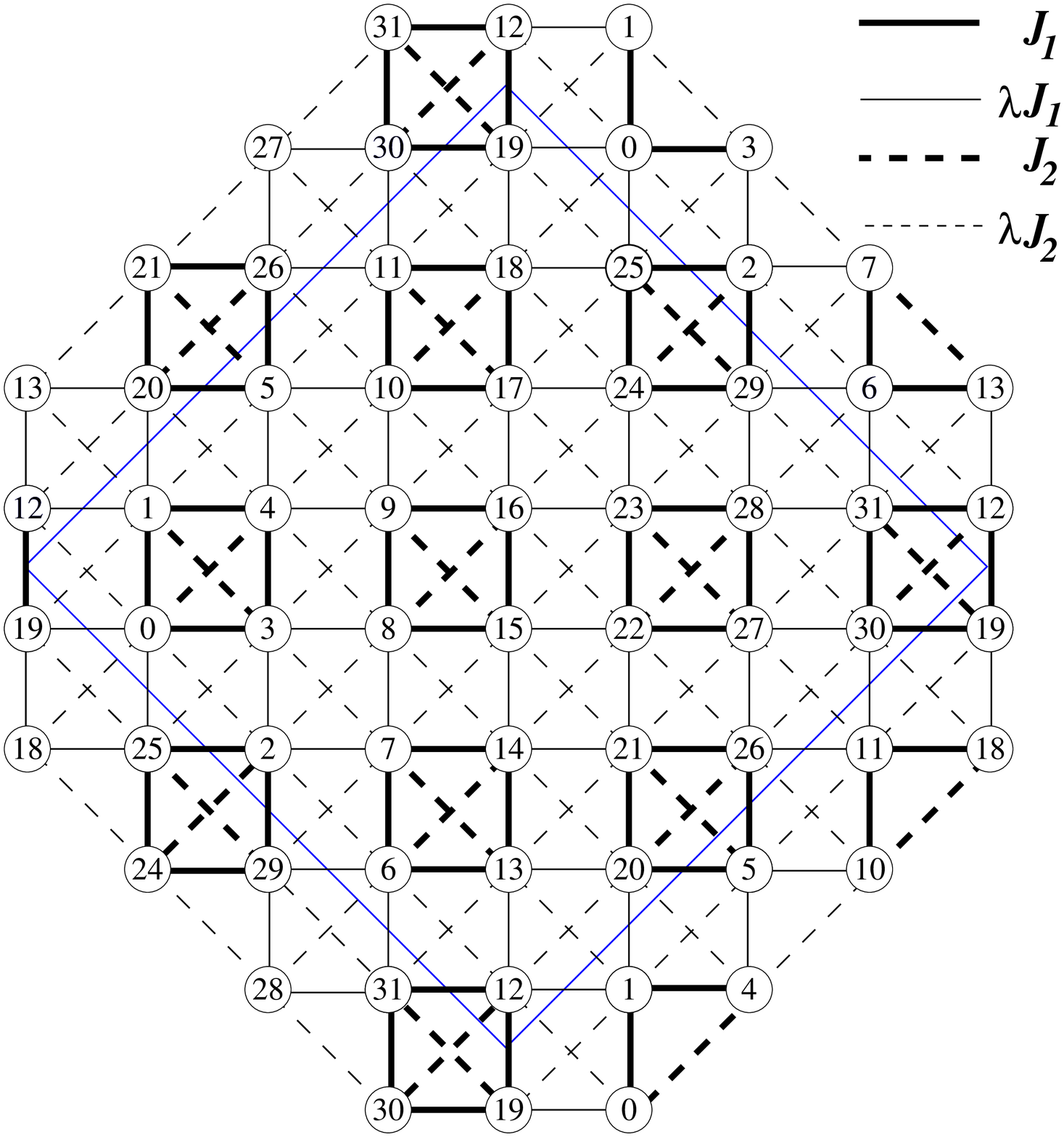,scale=0.3,angle=0.0}
\end{center}
\caption{The two-dimensional spin $1/2$ $J_1$-$J_2$ finite square lattice
of $N=32$ sites (periodic boundary conditions) with  plaquette structure
         (see Eq.~(\ref{ham})); 
         solid lines - NN bonds $J_1$ (thick) and $\lambda J_1$ (thin), dashed
         lines NNN bonds 
          $J_2$ (thick) and $\lambda J_2$ (thin).
}
\label{fig1}
\end{figure}

In this paper we consider the  $J_1$-$J_2$ model on the square lattice with plaquette structure
(see Fig.~\ref{fig1}) merging this way the two mechanisms to destroy magnetic
LRO, namely frustration and local singlet formation.
Moreover, the model we will consider corresponds to that proposed  in
Ref.~\onlinecite{ueda07} for (CuCl)LaNb$_2$O$_7$. 
The Hamiltonian of our model is given by
\be\label{ham}
 H = \sum_{a=0}^1 (\delta_{0a}+\delta_{1a}\lambda)\left(J_1\sum_{\langle
ij\rangle_a}\bfs_i\bfs_j + J_2\sum_{\langle\langle ij \rangle\rangle_a}\bfs_i\bfs_j\right)
,
\ee
where the sum is taken over the nearest ($\langle\cdots\rangle$) and next-nearest
($\langle\langle\cdots\rangle\rangle$) neighbors,  and $\delta_{ab}$ is the Kronecker
symbol. We consider FM and AFM NN bonds $J_1=\pm 1$ and AFM (i.e. frustrating) NNN
bonds $J_2 \ge 0$.  
The intra- ($a=0$) and inter-plaquette interaction  ($a=1$) differ by the
factor $\lambda$. 
The case $\lambda=1$ corresponds to
the standard uniform $J_1$-$J_2$ model on the square lattice, whereas the
limit $\lambda=0$ describes unconnected 4-spin $J_1$-$J_2$ pla\-quettes.
For the  parameter $\lambda$ we consider the interval $\lambda\in[0,1]$
this way interpolating between a system of isolated plaquettes and the
uniform $J_1$-$J_2$ model.

The model (\ref{ham}) has been studied previously by series expansion\cite{singh99} 
for AFM $J_1>0$, and by bond-operator mean-field theory as well
as a second-order perturbation theory in $\lambda$\cite{ueda07} for 
(FM) $J_1<0$. 
Bearing in mind the intensive work on the standard $J_1$-$J_2$ model, see
Refs.~\onlinecite{chandra88,dagotto89,schulz,richter93,richter94,
zhito96,Trumper97,bishop98,singh99,sushkov01,capriotti01,Sir:2006,Schm:2006,darradi08,
sousa,bishop08,xxz,ortiz,singh2009,cirac2009,ED40,fprg,cirac2010,balents2011,verstrate2011}
and references therein, it seems to
be desirable to discuss the much less studied plaquette $J_1$-$J_2$ model by
methods being alternative to those used in Refs.~\onlinecite{singh99}
and \onlinecite{ueda07}
this way to confirm or to question the findings of those papers.   
 
In this paper we consider both, the AFM ($J_1>0$) and the FM ($J_1<0$) 
version of the model (\ref{ham}), and we  derive the GS quantum phase diagram for
both versions of the model. We use the Lanczos exact diagonalization, see
Sec.~\ref{sec:ed}, and 
the coupled cluster method (CCM), see Sec.~\ref{sec:ccm}, 
to analyze the GS of the model.
Both methods  are powerful general many-body methods and, particularly, the
combination of both  
can lead to a consistent description of various 
aspects of quantum spin systems. We note that another important method for
quantum spin systems, the quantum Monte Carlo method (QMC), cannot be used for the
considered frustrated model,
since it suffers from the minus sign problem. 

\begin{figure*}[t]
\begin{center}
\epsfig{file=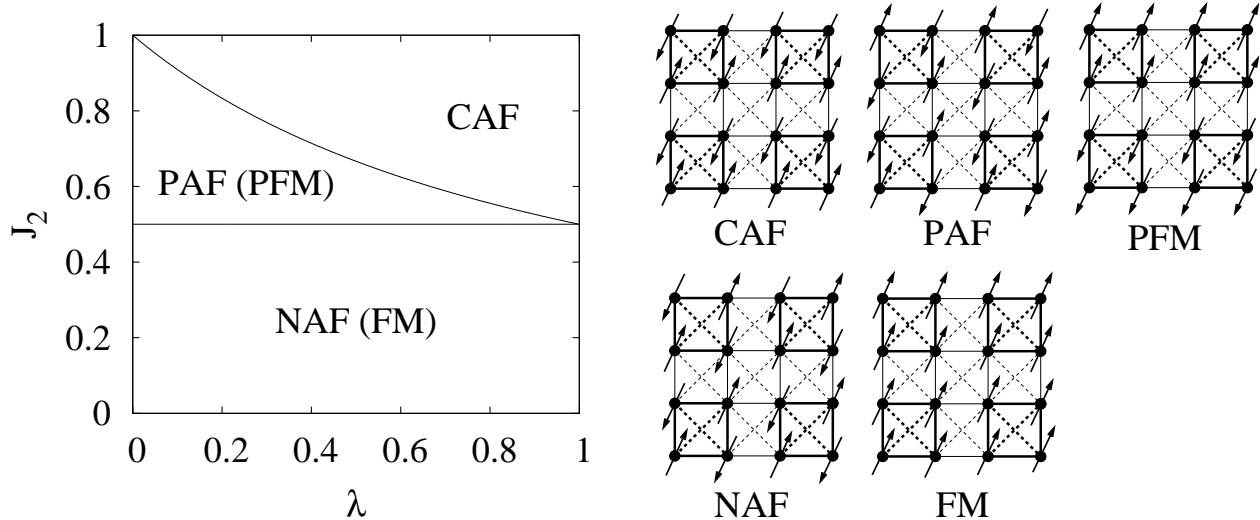,scale=0.35,angle=0.0}
\end{center}
\caption{Left panel: The classical phase diagram of the two-dimensional $J_1$-$J_2$
square-lattice Heisenberg model with 
plaquette structure, see Eq.~(\ref{ham}). For the AFM model ($J_1=+1$) 
the GS phases CAF, NAF, and PAF exist. For the FM model ($J_1=-1$)
the CAF GS phase also exists, but the NAF and the PAF phases are replaced by
the FM and the  PFM phases, respectively.
Right panel: Illustration of the classical
phases (CAF: collinear striped
antiferromagnetic  order, PAF: plaquette antiferromagnetic order, NAF: \Neel\ antiferromagnetic
order,
FM:  ferromagnetic order, PFM: plaquette ferromagnetic  order).}
\label{phadia_cl}
\end{figure*}

\section{Classical ground-state phases} \label{clasGS}
It is well-known that the GS for the classical case ($s\rightarrow\infty$) of the uniform
$J_1$-$J_2$ model (i.e., $\lambda=1$) has three phases: The NAF phase for
$J_2<J_1/2$ and $J_1>0$, the FM phase for $J_2<-J_1/2$ and $J_1< 0$, and
a phase consisting of two interpenetrating
\Neel-ordered square lattices. The angle between the directions  of the two
interpenetrating \Neel\; states is
arbitrary in the classical limit, whereas 
due to the {\it order-from-disorder} effect collinear
states are favored by fluctuations,
i.e., the CAF is realized.
For $\lambda < 1$ to the best of our knowledge the classical phase diagram was not
discussed so far.
By contrast to the uniform model, two additional plaquette phases appear for $\lambda
< 1$, see
Fig.~\ref{phadia_cl},
namely the plaquette antiferromagnetic  (PAF) phase separating the NAF from the CAF phase
for $J_1>0$ and
the plaquette ferromagnetic  (PFM) between the FM and the CAF phases
for $J_1<0$.
These plaquette phases have some relation to the CAF phase:
They also can be understood  as a system of two interpenetrating
\Neel-ordered square lattices, where again the angle between the directions  of the two
interpenetrating \Neel\; states is
arbitrary in the classical limit. 
However, as an elementary unit (a 'lattice site') of the two interpenetrating square lattices
does not act a single spin (site), but rather a 4-spin-(4-site-)plaquette
carrying strong bonds.
And again,
due to the {\it order-from-disorder} effect collinear
states are favored by fluctuations,
i.e., the collinear PAF or PFM phases are realized, respectively.
A graphical illustration of these classical phases is given in
Fig.~\ref{phadia_cl}.
The energy of the classical plaquette phases reads
$E_{\rm PAF, PFM}=-\left[\pm\frac{1}{4}J_1 - \frac{1}{8}J_2 \right ]
-\frac{1}{8}\lambda J_2$, where the upper (lower) sign corresponds to PAF
(PFM). Obviously,   
the inter-plaquette interaction is due to $\lambda J_2$ only.
The transition line between
the CAF and the PAF (PFM) phases is given by 
$J_2=J_1/(1+\lambda)$ ($J_2=-J_1/(1+\lambda)$), the transition between the NAF and the PAF
phases as well
as the FM and the PFM phases are horizontal lines $J_2=J_1/2$ and
$J_2=-J_1/2$, respectively.
Except the trivial degeneracies, the CAF, the PAF and the PFM states are two-fold
degenerate, since the stripes of parallel spins (CAF) or parallel plaquettes
(PAF, PFM) can be arranged either along the horizontal or the vertical
direction.

\section{Methods}
\subsection{Exact diagonalization of finite lattices}
\label{sec:ed}

The Lanczos exact diagonalization method was successfully used to discuss the
GS phases of the uniform $s=1/2$ 
$J_1$-$J_2$ model (i.e.,
$\lambda=1$) using finite lattices of $N=16,20,32,36$ and $40$ 
sites.\cite{dagotto89,schulz,richter93,richter94,ED40,richter10}
However,  the new classical phases, PAF and PFM, appearing due to the plaquette structure do
not
fit to the periodic boundary conditions of the finite lattices of $N=20,36$
and $40$ sites, and, therefore, we do not have the possibility to perform a finite-size extrapolation   
used previously for the uniform
model.\cite{schulz,ED40,richter10}
Using J\"org Schulenburg's
{\it spinpack}\cite{spinpack} we have calculated the GS focusing on the
finite lattice of  
$N=32$  sites shown in Fig.~\ref{fig1} for a large set of $\lambda$ and $J_2$
values fixing $J_1$ either to
$+1$ or to $-1$.

To analyze the GS magnetic ordering we have calculated the spin-spin
correlation function $\langle {\bf s}_0{\bf s}_i\rangle$ as well as a general
finite-size order
parameter\cite{wir04,bounce}
\begin{equation} \label{order_para}
  m^2
     =\frac{1}{(N-4)^2}
     \sum_{i,j \notin P }^{N-4}  {}|\langle{\bf s}_{i}{\bf s}_{j}\rangle|
     \, 
\end{equation}
that adds the total strength of the spin-spin
correlation functions.
Bearing in mind the strong intra-plaquette correlation functions for
$\lambda <1$ we have excluded these correlations in the sum, this way
increasing the weight of distant correlation functions.

\subsection{Coupled Cluster Method}
\label{sec:ccm}
We first mention that the coupled cluster method (CCM) yields results directly in
the thermodynamic limit $N\to\infty$.
The CCM has been previously reviewed in several
articles (e.g., for the AFM\cite{bishop98,bishop08,xxz,darradi08} and for
the
FM $J_1$-$J_2$ model\cite{richter10,li2012} as well as for dimerized
models,\cite{krueger00,krueger01,farnell05,farnell09})
and will not be repeated here in detail.
For more general information on the methodology of the CCM, see, e.g.,
Refs.~\onlinecite{zeng98,bishop98a,bishop00,bishop04}.

The CCM is a quantum many-body method and is defined by a reference (or model) state 
$|\Phi\rangle$ and a complete set of mutually commuting many-body creation operators $C_I^+$. 
For our model we choose the classical GSs, see Fig.~\ref{phadia_cl}, as reference
states. It is convenient to perform an appropriate 
rotation of the local axis of the spins such that in the rotated coordinate
frame the reference state is  a tensor product of spin down states
$|\Phi\rangle = |\hspace{-3pt}\downarrow\rangle |\hspace{-3pt}\downarrow\rangle |\hspace{-3pt}\downarrow\rangle\dots$.
The creation operators are then the multispin creation operators
$C_I^+ = s_i^+,\,\,s_i^+s_j^+,\,\,s_i^+s_j^+s_k^+,\cdots$ where the indices $i,j,k,\dots$ 
denote arbitrary lattice sites.

The CCM parameterizations of the ket- and bra- GSs are given by
(with $C_I^-=(C_I^+)^+$)
\begin{eqnarray}
\label{ccm}
|\Psi\rangle = e^S|\Phi\rangle, 
\qquad 
S = \sum_{I \neq 0}{\cal S}_IC_I^+ ; 
\nonumber\\
\langle\tilde{\Psi}| =  \langle\Phi|\tilde{S}e^{-S},
\qquad 
\tilde{S} = 1 + \sum_{I \neq 0}\tilde{\cal S}_IC_I^- .
\end{eqnarray}
Using $\la\Phi|C_I^+=0=C_I|\Phi\ra$ $\forall I\neq 0$, $C_0^+\equiv 1$,
the orthonormality condition $\la\Phi|C_IC_J^+|\Phi\ra=\delta_{IJ}$,
and the completeness relation $\displaystyle\sum_I C_I^+|\Phi\ra\la\Phi|C_I=1=|\Phi\ra\la\Phi|+\sum_{I\neq 0}C_I^+|\Phi\ra\la\Phi|C_I$
we get a set of non-linear and linear equations for 
the correlation coefficients ${\cal S}_I$ and $\tilde{\cal S}_I$, respectively.
The order parameter (sublattice magnetization) in the rotated coordinate frame
is given by  
\begin{equation}\label{CCM_order}
M = -\frac{1}{N} \sum_i^N \la\tilde\Psi|s_i^z|\Psi\ra.
\end{equation}
In the CCM the only approximation is the truncation of the expansion of 
the correlation operator $S$ and $\tilde S$. We use the well established
LSUB$m$
scheme, where
all multispin correlations on the lattice with $m$ or fewer contiguous sites
are taken into account.
The number of these configurations can be reduced using 
lattice symmetry and conservation laws, but it is increasing very rapidly with
$m$.
In the highest order of approximation considered here, LSUB10, we have, for example, 
180957 configurations for the collinear stripe reference state (CAF) and 219446
for the antiferromagnetic  plaquette  state (PAF), i.e., finally 
180957 or  219446
corresponding coupled non-linear equations which have to be solved numerically.

Since the LSUB$m$ approximation scheme becomes exact for $m \to \infty$, we can improve our results by 
extrapolating the ``raw'' LSUB$m$ data to $m \to \infty$. 
There is ample empirical experience regarding how one should
extrapolate 
the magnetic order parameter $M(m)$ 
for systems with quantum phase transition between magnetically ordered and
disordered GS phases.
Following Refs.~\onlinecite{bishop08,xxz,darradi08,richter10} we use
$M(m)=b_0+b_1(1/m)^{1/2}+b_2(1/m)^{3/2}$ to extrapolate to $m \to \infty$.  
As a rule\cite{bishop08,xxz,darradi08,richter10} the lowest level of approximation, 
LSUB2, is excluded from extrapolation. For the particular 
4-spin plaquette model considered here, we expect LSUB2 to be especially
poor, because its cluster size (2 spins) is smaller than the unit cell 
(4 spins) of the model.
Hence, we use LSUB4, LSUB6, LSUB8 and LSUB10 data for the extrapolations.

\begin{figure}[ht]
\begin{center}
\epsfig{file=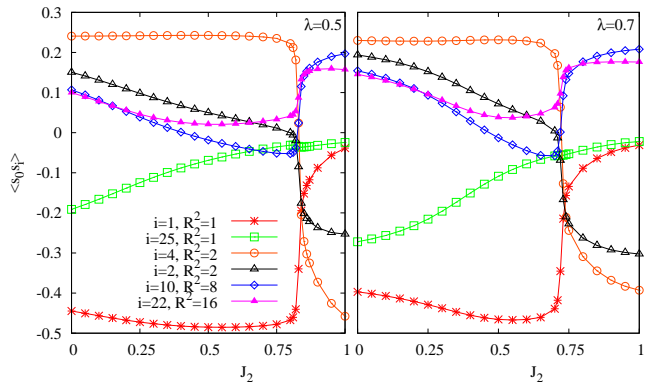,scale=0.55,angle=0.0}
\end{center}
\caption{ED data for the spin-spin correlation functions $\langle {\bf s}_0{\bf s}_i\rangle$
versus $J_2$ (AFM $J_1=+1$) for two values of
$\lambda$ for the finite lattice of $N=32$ sites.
Except the site indeces $0$ and $i$ corresponding to  
Fig.~\ref{fig1} we also give the square of separation $R^2$ of the sites $0$
and $i$. Note that $\langle {\bf s}_0{\bf s}_1\rangle$ and $\langle {\bf s}_0{\bf
s}_4\rangle$ are intra-plaquette correlation functions, whereas the other
ones are inter-plaquette correlation functions. }
\label{figsisjAFM}
\end{figure}

\begin{figure}[ht]
\begin{center}
\epsfig{file=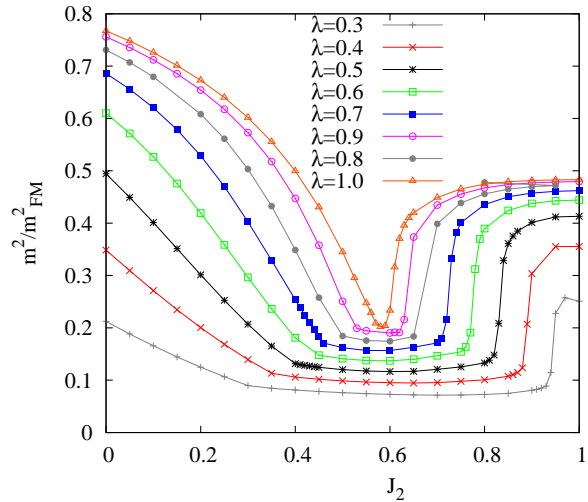,scale=0.8,angle=0.0}
\end{center}
\caption{ED data for the finite-size order parameter $m^2/m^2_{FM}$ as
defined in
Eq.~(\ref{order_para})
versus $J_2$ (AFM $J_1=+1$) for various  values of
$\lambda$ for the finite lattice of $N=32$ sites.
For convenience $m^2$ is scaled by its value for the ferromagnetic state
$m^2_{FM}=0.25$.
}  
\label{order_ED_AFM}
\end{figure}

\section{Results for the quantum $s=1/2$ model }
\label{sec:res}
\subsection{Antiferromagnetic nearest-neighbor exchange $J_1$}
\label{sec:res_a}
We first consider the AFM case, set $J_1=+1$, and start with the
discussion of ED data, see Figs.~\ref{figsisjAFM} and \ref{order_ED_AFM}.
Corresponding to the three classical phases there are three regimes in the
quantum model. 
The spin-spin correlation functions $\langle {\bf s}_0{\bf s}_i\rangle$ 
presented in Fig.~\ref{figsisjAFM} for
$\lambda=0.5$ and $\lambda=0.7$ are quite strong for small and large
$J_2$ thus indicating semi-classical magnetic LRO, where the signs  of
$\langle {\bf s}_0{\bf s}_i\rangle$
 fit to the classical phases NAF and CAF.
In an intermediate regime, around $J_2=0.5$, the inter-plaquette spin-spin correlations are
weak.
These three different regimes are also well seen in Fig.~\ref{order_ED_AFM},
where the finite-size order parameter $m^2$, see Eq.~(\ref{order_para}), is shown.
In particular, the $m^2$ data yield clear evidence for regions with weak
magnetic order. The widths of those regions increase drastically with
decreasing $\lambda$.
Bearing in mind that for $\lambda=1$ the region of magnetic disorder is $0.4
\lesssim J_2 \lesssim 0.6$, see, e.g.,
Refs.~\onlinecite{Sir:2006,darradi08,ortiz,fprg,ED40}, and \onlinecite{balents2011}, our ED data for $N=32$ suggest that there is no
semi-classical
antiferromagnetic LRO of plaquette type (PAF).
Moreover, for small values of $\lambda \lesssim 0.5$ the finite-size order
parameter is small in the whole range of $J_2$ values indicating the absence
of any magnetic LRO.

\begin{figure}[ht]
\begin{center}
\epsfig{file=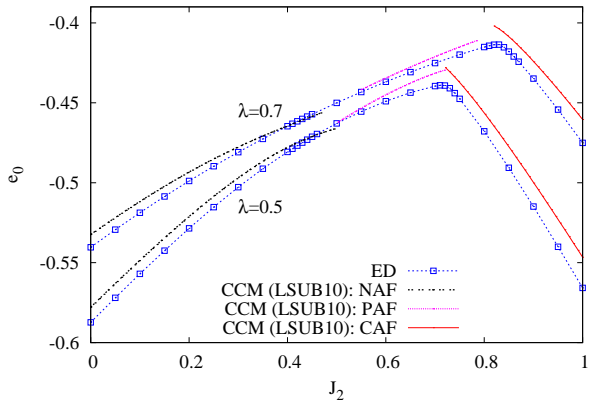,scale=0.65,angle=0.0}
\end{center}
\caption{CCM LSUB10 data for the GS energy $e_0$ 
         (AFM $J_1=+1$) for $\lambda=0.5$ and
         $\lambda=0.7$
          compared with corresponding ED data ($N=32$).
         The CCM results correspond to the NAF reference state (small
         $J_2$), the PAF reference state (intermediate $J_2$), and the CAF
         reference state (large $J_2$).
}
\label{e_CCM_AFM}
\end{figure}

\begin{figure}[ht]
\begin{center}
\epsfig{file=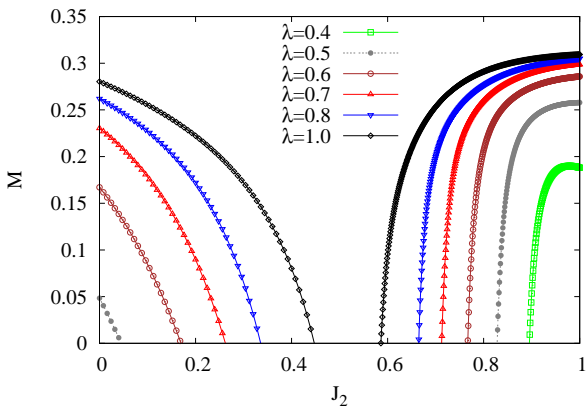,scale=0.65,angle=0.0}
\end{center}
\caption{Extrapolated CCM data for the magnetic order parameter (sublattice
magnetization) $M$ for the AFM case 
         ($J_1=+1$) and for various values of $\lambda$. 
        The CCM reference state  is the NAF state (left hand side, small  $J_2$) and the CAF state 
        (right hand side,  large  $J_2$).
         The extrapolated CCM data are obtained using LSUB$m$ with
          $m=4,6,8,10$ and
         the extrapolation rule $M(m)=b_0+b_1(1/m)^{1/2}+b_2(1/m)^{3/2}$.
}
\label{fig3}
\end{figure}

\begin{figure}[ht]
\begin{center}
\epsfig{file=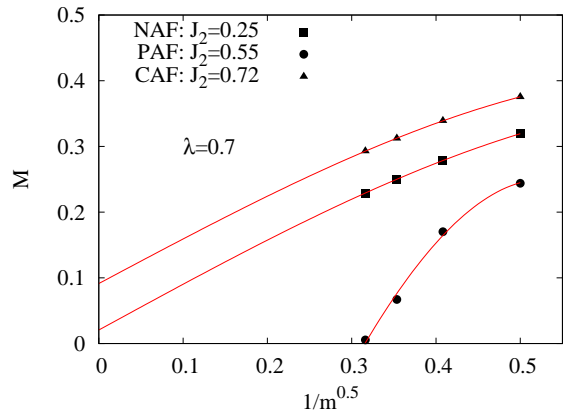,scale=0.7,angle=0.0}
\end{center}
\caption{Illustration of the extrapolation (solid lines) of the CCM-LSUB$m$ data (symbols)
for
 the magnetic order parameter $M$
for 
$\lambda=0.7$, $J_1=+1$ and specific values of $J_2$ corresponding to the NAF, PAF,
CAF reference states.
For the extrapolations to $m \to \infty$  according to
$M(m)=b_0+b_1(1/m)^{1/2}+b_2(1/m)^{3/2}$
we have used  LSUB$m$ data for $m=4,6,8,10$.
}
\label{fit_AFM}
\end{figure}

Now we discuss the CCM results ($N \to \infty$).
Fig.~\ref{e_CCM_AFM} provides the CCM GS energy per site $e_0$ for $\lambda =
0.5$ and $\lambda = 0.7$
compared with the corresponding ED data.
Obviously, the CCM and the ED agree well. 
To discuss magnetic LRO we consider the CCM-LSUB$m$ magnetic order parameter
$M$ (sublattice magnetization)
extrapolated to $m\to\infty$, see Eq.~(\ref{CCM_order}), that is depicted in
Fig.~\ref{fig3}.
To illustrate the quality of the used extrapolation for the order parameter
$M$ of the
`raw' LSUB$m$ data to the limit $m \to \infty$ we also show, as an example, corresponding plots
for the $\lambda=0.7$
in Fig.~\ref{fit_AFM}. It is obvious that the  LSUB$m$ data are well fitted by
the applied extrapolation function.

\begin{figure}[ht]
\begin{center}
\epsfig{file=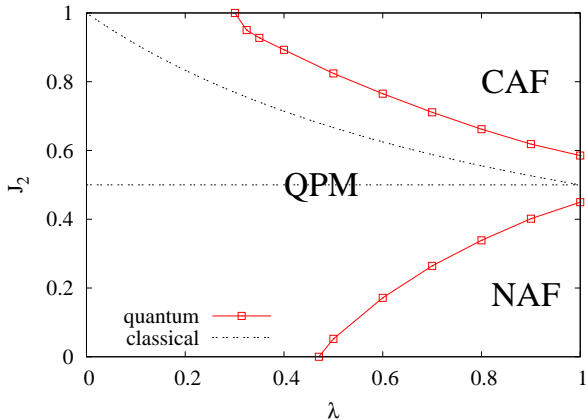,scale=0.65,angle=0.0}
\end{center}
\caption{The CCM GS phase diagram of the model (\ref{ham}) for the AFM case 
         ($J_1=+1$). NAF denotes the semi-classical phase with \Neel~LRO, 
CAF the semi-classical phase with collinear
striped antiferromagnetic LRO, and QPM the magnetically disordered
quantum paramagnetic phase.  For comparison we also show the classical
transition lines (thin dashed lines), cf. Fig.~\ref{phadia_cl}.
}
\label{fig4}
\end{figure}

First we notice that the extrapolated order parameter using the PAF reference
state for all relevant parameter sets of $\lambda$ and $J_2$ vanishes (see
also  Fig.~\ref{fit_AFM}), i.e., there is no PAF magnetic LRO in the quantum
model. This finding is in agreement with the ED data for $m^2$ discussed above
(and see again Fig.~\ref{order_ED_AFM}).  
The range of stability of the semi-classical NAF and CAF phases is visible in 
Fig.~\ref{fig3}.
We define the quantum critical points $J_2^{c1}(\lambda)$ and
$J_2^{c2}(\lambda)$
 as those points, where the extrapolated CCM magnetic order parameter $M$
vanishes. For $\lambda=1$ (uniform model) we find $J_2^{c1} = 0.447J_1$ and
$J_2^{c2} = 0.586J_1$ which is in agreement with previous CCM
predictions.\cite{bishop08,xxz,darradi08} As expected, with decreasing of $\lambda$ the
region of $J_2$ values without magnetic LRO increases,
see Fig.~\ref{fig3}. 
Collecting the data for the quantum critical points
$J_2^{c1}$ and $J_2^{c2}$ for various $\lambda$ we get the GS phase diagram
of the quantum model as
shown Fig.~\ref{fig4}. Our phase diagram is in good agreement with the only
available corresponding one of Ref.~\onlinecite{singh99}. Hence, one can
argue that the phase diagram is basically correct.

Concerning the order of the quantum phase transitions at
$J_2^{c1}$ and $J_2^{c2}$ our data are in favor of a continuous transition at
$J_2^{c1}$ and a first-order transition at $J_2^{c2}$ as it is was discussed 
for the uniform model.\cite{schulz,singh99,sushkov01,ED40}  
The scenario of a first-order transition between the QPM and the CAF phases
is supported by (i)
the kink-like behavior in $e_0$, see Fig.~\ref{e_CCM_AFM}, (ii)
the
steep fall in the CCM order parameter $M$ near  $J_2^{c2}$, see
Fig.~\ref{fig3},
and (iii) the
jump-like behavior of the spin-spin
correlations functions, see Fig.~\ref{figsisjAFM},  and of the finite-size order
parameter, see Fig.~\ref{order_ED_AFM}.

For the unfrustrated case $J_2=0$ we can compare our result $\lambda_c=0.47$
(and see Fig.~\ref{fig4}) for the critical
$\lambda$ where the  NAF LRO breaks down with several  previous
results, namely 
 $\lambda_c=0.555$ (Ising series expansion\cite{singh99})
$\lambda_c=0.6$ (exact diagonalization\cite{voigt02})
$\lambda_c=0.5485$ (QMC\cite{fabricio08,wenzel09}),
 $\lambda_c=0.5491\dots0.5513$ (contractor renormalization
expansion\cite{fabricio08}),
and  $\lambda_c=0.4822$ (real space renormalization group approach\cite{fledderjohann09}). 
Likely, the QMC result is most accurate. 
Our result is in reasonable  agreement with that result, but slightly
overestimates the stability region of NAF LRO.

Let us briefly discuss another limit of the model, namely the limit of large
$J_2$. As discussed above in that limit the system splits into two interpenetrating
square-lattice Heisenberg antiferromagnets (where  $J_2$ plays the role of the AFM NN
bond), 
which are \Neel-ordered for $\lambda=1$. Hovewer,
for $\lambda < 1$  each of the two interpenetrating
square lattices carries now a staggered arrangement of $J_2$ and $\lambda J_2$ bonds,
which corresponds precisely to the so-called $J$-$J'$ model discussed in
Refs.~\onlinecite{singh88,ivanov96,krueger00,wenzel08,wenzel09,vidal09}. The
QMC estimate of the critical $\lambda$ is $\lambda_c=0.397$, see
Ref.~\onlinecite{wenzel08}. 
For $J_2=1$  we find from our CCM data a critical value of
$\lambda_c=0.301$ (see
Fig.~\ref{fig4}). Increasing of $J_2$ beyond  $J_2=1$ (not shown in
Fig.~\ref{fig4}) yields a steep increase of the  critical line as already
indicated by the 
last two data points near $J_2=1$ in Fig.~\ref{fig4}. In the limit of large
$J_2$ the critical
$\lambda_c$ is even smaller as  for $J_2=1$.
For example at $J_2=5$ we obtain  
 $\lambda_c= 0.375$ that is in good agreement with the QMC result for
the critical $\lambda_c$ of the $J$-$J'$ model. 
Note that our value of $\lambda_c$ deviates from an early  CCM result 
$\lambda_c=0.316$.\cite{krueger00} The
difference is related to the fact that we use here (i) a higher 
approximation level (LSUB10) and (ii) an improved extrapolation in comparison
with Ref.~\onlinecite{krueger00}.

\subsection{Ferromagnetic nearest-neighbor exchange $J_1$}
\label{sec:res_b}

\begin{figure}[ht]
\begin{center}
\epsfig{file=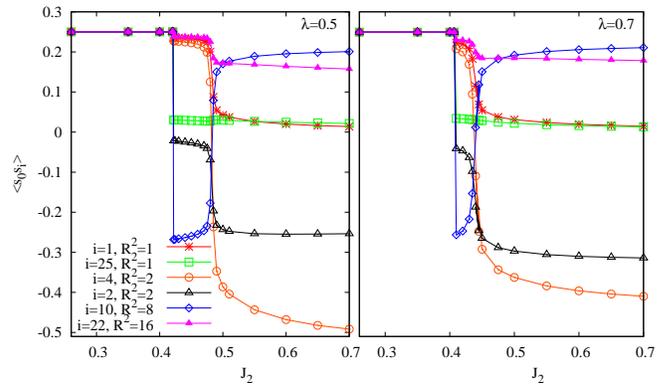,scale=0.55,angle=0.0}
\end{center}
\caption{ED data for the spin-spin correlation functions $\langle {\bf s}_0{\bf
s}_i\rangle$
versus $J_2$ (FM $J_1=-1$) for two values of
$\lambda$ for the finite lattice of $N=32$ sites.
Except the site indeces $0$ and $i$ corresponding to
Fig.~\ref{fig1} we also give the square of separation $R^2$ of the sites $0$
and $i$. Note that $\langle {\bf s}_0{\bf s}_1\rangle$ and $\langle {\bf
s}_0{\bf
s}_4\rangle$ are intra-plaquette correlation functions, whereas the other
ones are inter-plaquette correlation functions.
}
\label{figsisjFM}
\end{figure}

\begin{figure}[ht]
\begin{center}
\epsfig{file=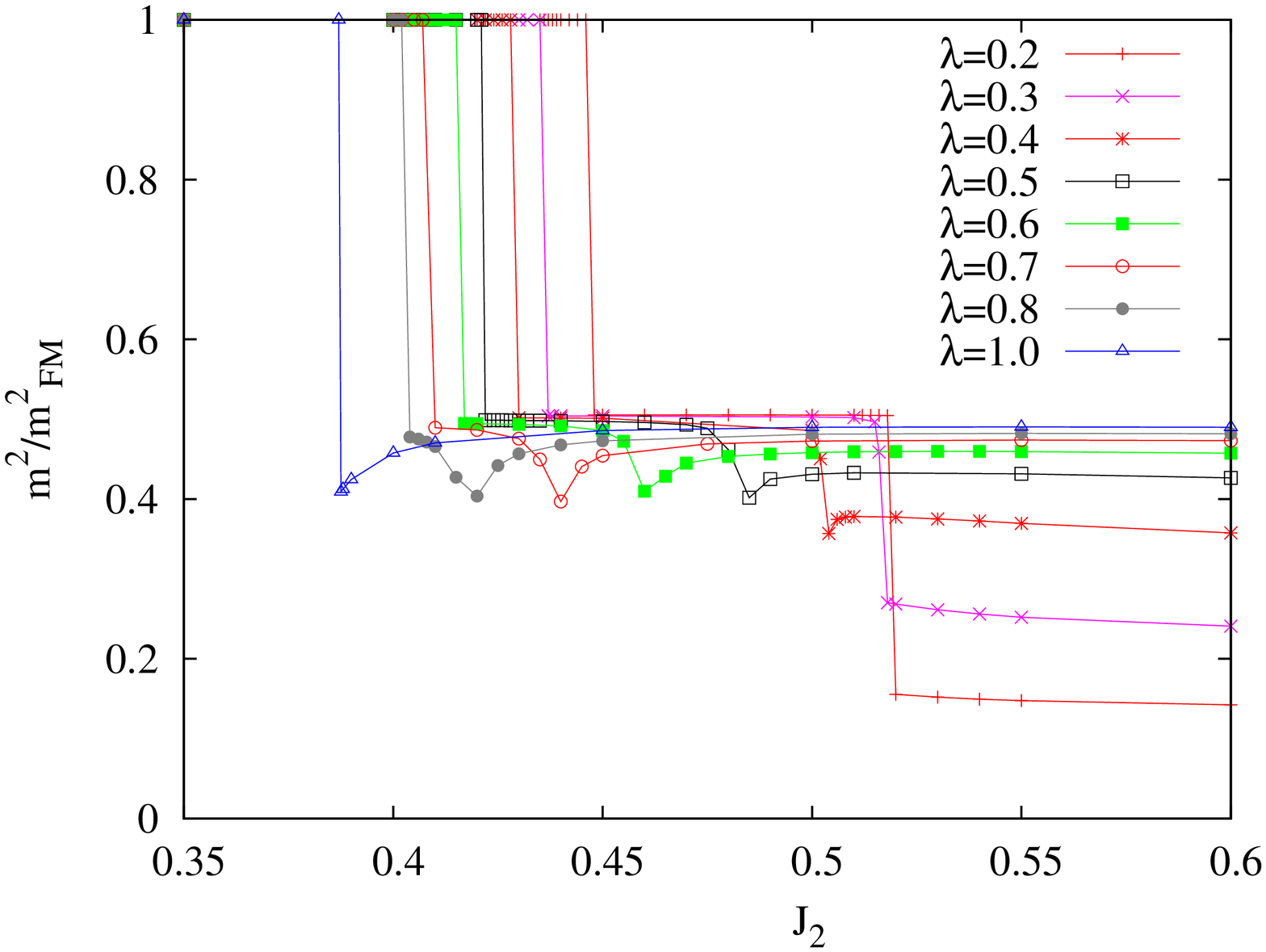,scale=0.35,angle=270.0}
\end{center}
\caption{ED data for the finite-size order parameter $m^2/m^2_{FM}$ as defined in
Eq.~(\ref{order_para})
versus $J_2$ (FM $J_1=-1$) for various  values of
$\lambda$ for the finite lattice of $N=32$ sites. 
For convenience $m^2$ is scaled by its value for the ferromagnetic state $m^2_{FM}=0.25$.
}
\label{order_ED_FM}
\end{figure}
\begin{figure}[ht]
\begin{center}
\epsfig{file=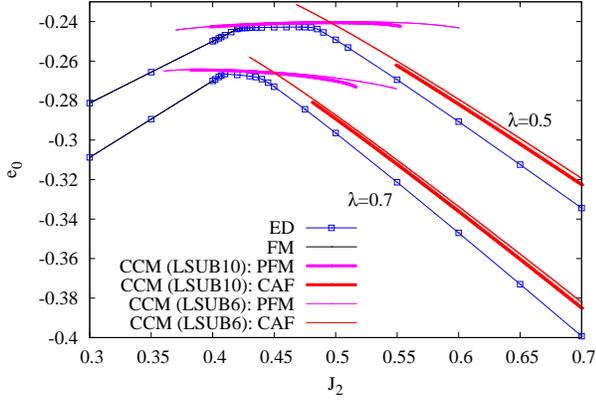,scale=0.65,angle=0.0}
\end{center}
\caption{CCM-LSUB10 (thick lines) and LSUB6 data (thin lines) for the GS energy $e_0$ 
         ($J_1=-1$) for $\lambda=0.5$ and
         $\lambda=0.7$
          compared with corresponding ED data ($N=32$).
         The CCM results correspond to 
          the PFM reference state (intermediate $J_2$) and the CAF
         reference state (large $J_2$). The black solid line is the exact FM
         GS energy. 
}
\label{e_CCM_FM}
\end{figure}

We now consider the FM case and set $J_1=-1$.
Although, the classical phase diagrams for $J_1=-1$ and $J_1=+1$ are identical,
see Fig.~\ref{phadia_cl}, we know from the uniform model (i.e., $\lambda=1$)
that the phase diagrams in the quantum case $s=1/2$ are basically
different.\cite{shannon04,shannon06,sindz07,schmidt07,schmidt07_2,sousa,shannon09,
sindz09,haertel10,richter10,momoi2011,cabra2011} In particular, it was
found for $J_1=-1$ that the region of semi-classical CAF LRO extends up to much smaller
values of $J_2$ compared with  the case of $J_1=+1$.\cite{richter10,cabra2011} Hence, we expect
also for $\lambda < 1$ basic differences between $J_1=+1$ and $J_1=-1$.
   
We start again with ED results for the
 spin-spin correlation functions $\langle {\bf s}_0{\bf s}_i\rangle$, see
Fig.~\ref{figsisjFM}, and 
the finite-size order parameter $m^2$ defined in Eq.~(\ref{order_para}), see
Fig.~\ref{order_ED_FM}.
The three classical phases lead again to three different regimes in the
quantum model. 
The trivial FM state at small $J_2$ is of course also present in the quantum model.
It gives way  at $J_2^{c1}(\lambda)$
for an intermediate singlet state. This transition at $J_2^{c1}$ is clearly seen by 
jumps in  $\langle {\bf s}_0{\bf s}_i\rangle$ and    $m^2$. 
$J_2^{c1}$ is smaller than
the classical value $J_{2,clas}^{c1}=0.5$, and it
depends on $\lambda$ for the quantum model.
Moreover, there is a second jump-like behavior of the spin-spin correlation
functions indicating the transition from the intermediate regime to the CAF
regime.
We can use the positions of these jumps to extract the transition points
$J_2^{c1}$ and $J_2^{c2}$
between the regimes from our ED calculations, see the discussion 
of Fig.~\ref{fig6} given below.    

The main difference in comparison to the AFM model ($J_1=+1$) concerns 
the intermediate regime. From Fig.~\ref{figsisjFM} it is evident that
the inter-plaquette correlation functions  $\langle {\bf s}_0{\bf s}_{10}\rangle$ and $\langle {\bf s}_0{\bf s}_{22}\rangle$, see
Fig.~\ref{fig1}, are not small.
Hence, the intermediate phase might be long-range ordered for FM $J_1$.         
Indeed, in the finite-size order parameter $m^2$ shown in
Fig.~\ref{order_ED_FM} an intermediate regime is clearly visible for $\lambda
\lesssim 0.5$, where the finite-size order parameter $m^2$ in this regime
is  much larger than that for
$J_1=+1$ (cf. Fig.~\ref{order_ED_AFM}). 
The possible appearance of magnetic LRO can be understood on the basis of the classical PFM state (see
also Sec.~\ref{clasGS}): The 
elementary unit is the 4-spin-plaquette
(with strong bonds).  Due to the strong FM intra-plaquette bond $J_1$ the
plaquette  carries an
effective  block (composite) spin $s=2$. These block spins interact via
$\lambda J_2$ and form effectively two
interpenetrating square-lattice $s=2$ Heisenberg antiferromagnets.

\begin{figure}[ht]
\begin{center}
\epsfig{file=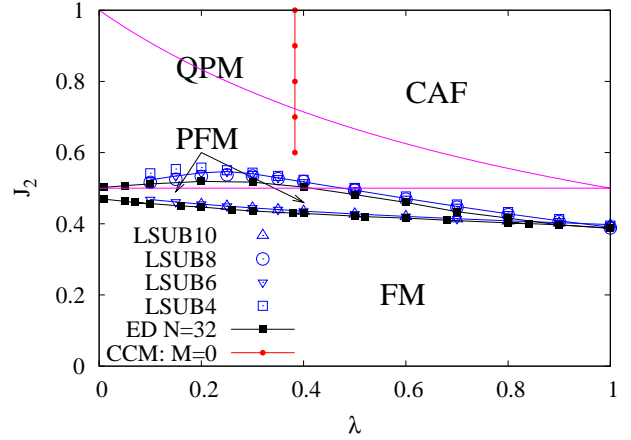,scale=0.68,angle=0.0}
\end{center}
\caption{GS phase diagram of the plaquette model of Eq.~(\ref{ham}) for the FM case 
         ($J_1=-1$). The transition lines between the FM and the PFM phases
as well as the PFM and the CAF/QPM  phases are obtained by (i) ED by using  the jumps in
the ED data (black lines), see Figs.~\ref{figsisjFM} and \ref{order_ED_FM}, and (ii) by
CCM from 
the intersection points between the GS energies using various reference states for the CCM
calculation (blue lines), see text.
The transition lines between the CAF and the QPM phase (red line) correspond
to those parameter values ($\lambda, J_2$), where the  
extrapolated CCM order parameter vanishes (and see Fig.~\ref{fig5}).
For comparison we also show the classical
transition lines (magenta line), cf. Fig.~\ref{phadia_cl}.
}
\label{fig6}
\end{figure}
\begin{figure}[ht]
\begin{center}
\epsfig{file=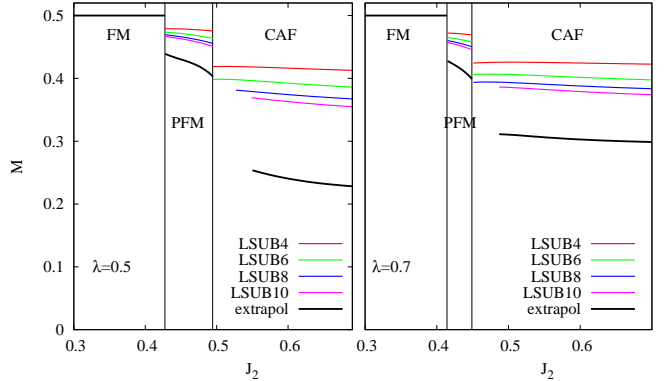,scale=0.56,angle=0.0}
\end{center}
\caption{CCM magnetic order parameter $M$ for the FM case 
         ($J_1=-1$) versus $J_2$ using the PFM and the  CAF states as CCM reference
          states for $\lambda=0.5$ and $\lambda=0.7$.
         The extrapolated CCM results are obtained using LSUB$m$ with $m=4,6,8,10$ and
         the extrapolation rule $M(m)=b_0+b_1(1/m)^{1/2}+b_2(1/m)^{3/2}$.
}
\label{FM_MvsJ2}
\end{figure}

\begin{figure}[ht]
\begin{center}
\epsfig{file=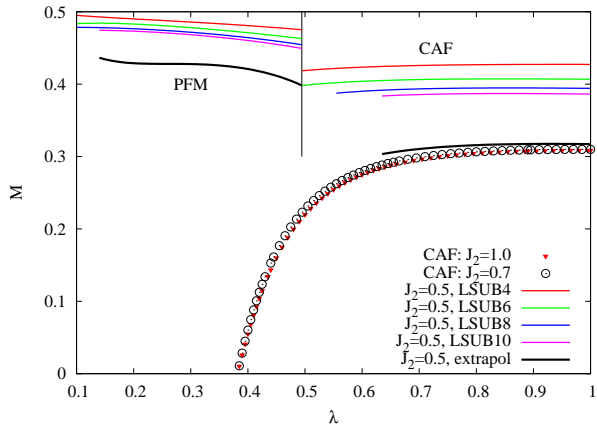,scale=0.65,angle=0.0}
\end{center}
\caption{Magnetic order parameter $M$ for the FM case 
         ($J_1=-1$) versus $\lambda$ for $J_2=1.0$ (filled triangles), $0.7$
        (circles) and $0.5$ (lines) using the PFM and the  CAF states as CCM reference
         states.
         The extrapolated CCM results are obtained using LSUB$m$ with $m=4,6,8,10$ and
         the extrapolation rule $M(m)=b_0+b_1(1/m)^{1/2}+b_2(1/m)^{3/2}$.         
}
\label{fig5}
\end{figure}

Now we pass to the CCM results ($N \to \infty$).
Fig.~\ref{e_CCM_FM} provides the CCM GS energy per site $e_0$ for $\lambda
= 0.5$ and $\lambda = 0.7$ in LSUB6 and LSUB10 approximation
compared with the corresponding ED data.
The agreement between the CCM and the ED is good.
The three regimes, FM, PFM, and CAF, are clearly seen in the energy
data.
For the CCM estimate for the transition point $J_2^{c1}$ between the FM state and the  PFM
regime we use the intersection points between the FM energy and the CCM-LSUB10
as well as the LSUB6
energy calculated with the PFM reference state. Both results for $J_2^{c1}$
almost coincide, see Fig.~\ref{fig6}. 
Moreover, we find an excellent agreement between the ED and CCM results for
$J_2^{c1}$.
Concerning the order of the phase transition at
$J_2^{c1}$ we have clear evidence  for a first-order transition as indicated
by 
(i) the
jump-like behavior of the spin-spin
correlations functions, see Fig.~\ref{figsisjFM},  and of the finite-size
order
parameter, see Fig.~\ref{order_ED_FM}, 
(ii)
the kink-like behavior in $e_0$, see Fig.~\ref{e_CCM_FM}, and (iii)
the
jump in the CCM order parameter $M$ at  $J_2^{c1}$, see
Fig.~\ref{FM_MvsJ2}.

To determine the transition point $J_2^{c2}$ between the PFM  and
the CAF regimes we are faced with the problem 
that  near this transition
no CCM-LSUB10 solutions for the CAF reference state could be found.
This observation, that for high orders of CCM approximation
in the vicinity of an intersection point of two CCM energy curves belonging
to two different GS phases 
no solution of
the large set of coupled nonlinear ket equations can be obtained, is often
found when dealing with strongly frustrated systems, see,
e.g., Refs.~\onlinecite{Schm:2006} and \onlinecite{richter10}.
However, from Fig.~\ref{e_CCM_FM}    
it is obvious that the parameter range, where solutions for lower levels $m$ of
LSUB$m$ approximations can be found, is
much larger than for LSUB10. Moreover, the LSUB$m$ data for $e_0$ for
various $m$ are very close to
each other. Hence, we can use the intersection points for lower levels
of CCM-LSUB$m$ approximation to determine the
second transition point $J_2^{c2}$. Thus, we find intersection points for
LSUB8, LSUB6, and  LSUB4 for $\lambda \ge 0.7$, $\lambda \ge 0.5$, and
$\lambda \ge 0.25$, respectively. 
In addition, we can also take benefit from the almost straight behavior of $e_0(J_2)$
curves, see
Fig.~\ref{e_CCM_FM}, which allows a reliable extrapolation of $e_0(J_2)$ until
a hypothetical intersection point, cf. also Refs.~\onlinecite{richter10} and
\onlinecite{cabra2011}.
This gives finally various sets of $J_2^{c2}(\lambda)$ data, which, however,
agree well with each other, cf. Fig.~\ref{fig6}. Only for $\lambda \lesssim 0.3$ a
slight difference is visible.
Nevertheless, it is necessary to mention that  the CCM
estimate for $J_2^{c2}(\lambda)$ becomes less reliable for smaller values of
$\lambda$
due to the increasing  distance between the hypothetical intersection points
and the last data points for which LSUB$m$ solutions for the CAF
reference state could be found.
\begin{figure}[ht]
\begin{center}
\epsfig{file=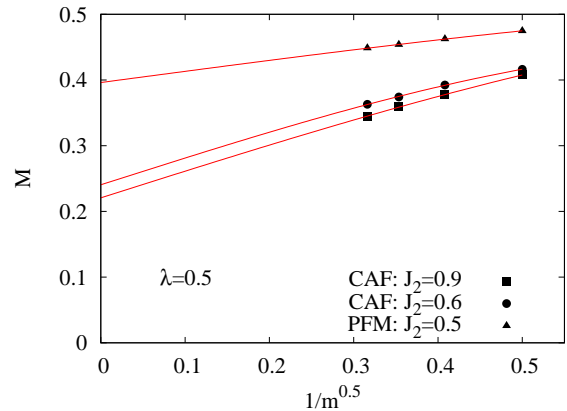,scale=0.7,angle=0.0}
\end{center}
\caption{Illustration of the extrapolation (solid lines) of the CCM-LSUB$m$ data (symbols)
for
 the magnetic order parameter $M$
for 
$\lambda=0.5$, $J_1=-1$  and specific values of $J_2$ corresponding to the
CAF and PFM
reference states.
For the extrapolations to $m \to \infty$  according to
$M(m)=b_0+b_1(1/m)^{1/2}+b_2(1/m)^{3/2}$
we have used  LSUB$m$ data for $m=4,6,8,10$.
}
\label{fit_FM}
\end{figure}
The results for $J_2^{c1}(\lambda)$and $J_2^{c2}(\lambda)$ obtained by ED
and CCM are collected in the phase diagram presented in
Fig.~\ref{fig6}. It is obvious that both, ED and CCM, yield very similar values
for $J_2^{c1}$ and $J_2^{c2}$.

The question about magnetic LRO in the various regimes we address next  
by analyzing the CCM magnetic order parameter $M$ for the PFM and CAF
regimes. (Remember that in the trivial FM phase we have always $M=1/2$.) 
In Fig.~\ref{FM_MvsJ2} we show $M$ versus $J_2$ for $\lambda = 0.5$ and
$\lambda = 0.7$. Obviously, the order parameter in the PFM regime is
non-zero and even larger than in the CAF regime. This observation is in
agreement with our ED results, cf. Fig.~ \ref{order_ED_FM}, and can
be related to the effective block-spin $s=2$ model discussed above.
The order parameter in the CAF is only weakly dependent on $J_2$, but its
magnitude depends on $\lambda$. As discussed above we do not get CAF CCM
solutions for LSUB$m$ for higher values of $m$ near the transition to the
PFM phase. Nevertheless, the curves presented in Fig.~\ref{FM_MvsJ2}
indicate that there is likely a direct first-order transition at $J_2^{c2}$ between
semi-classical phases with PFM and CAF
magnetic LRO for $\lambda \gtrsim 0.4$.

Next we fix $J_2$ and consider the behavior of the order parameter $M$ in
dependence on $\lambda$, see Fig.~\ref{fig5}.
For values of $J_2 \gtrsim 0.55$ the PFM regime is not relevant. 
As explained in Sec.~\ref{sec:res_a} in the limit of large $J_2$ our plaquette model
corresponds to the $J$-$J'$ model, where at $\lambda=\lambda_c=0.397$ (QMC
result, see
Ref.~\onlinecite{wenzel08})
a second-order transition to a QPM phase takes place.
This behavior is clearly seen in our CCM data for $M(\lambda)$ for
$J_2=1.0$ and $J_2=0.7$ shown in Fig.~\ref{fig5}. 
We find that for $J_2\gtrsim 0.6$ the extrapolated CCM order parameter $M$ vanishes
continuously at a
critical value $\lambda = \lambda_c(J_2)$ defining a second-order transition
between a semi-classical phase with CAF magnetic LRO and the QPM phase,
that is depicted as a red line in Fig.~\ref{fig6}.
Interestingly, $\lambda_c$ practically does not depend on $J_2$ and we have
$\lambda_c \approx 0.383$ for $0.6 \lesssim J_2 \le 1$. 
This value is close to the QMC value\cite{wenzel08} $\lambda_c=0.397$
valid for $J_2 \to \infty$ and also to $\lambda_c=0.375$ obtained for the AFM
model ($J_1=+1$) for $J_2=5$, see Sec.~\ref{sec:res_a}.

For  $J_2=0.5$ the situation is different, since both, the PFM and CAF
regimes, play a role. For $\lambda < 0.494$ we are inside the PFM regime.
Obviously, the PFM order parameter is large and the variation with
$\lambda$ is weak. Within the CAF regime
we are again faced with the
problem that the CAF LSUB$m$ solutions for higher $m$ terminate before meeting the
corresponding PFM LSUB$m$ solutions.
Hence, the critical line $ \lambda_c(J_2)$ (red vertical line in
Fig.~\ref{fig6}) terminates before meeting the transition line
$J_2^{c2}(\lambda)$, and we cannot give a secure statement on the
continuation of the critical line $ \lambda_c(J_2)$ towards lower values of
$J_2$.

Finally, we illustrate the quality of the used extrapolation for the order parameter
$M$ of the
`raw' LSUB$m$ data to the limit $m \to \infty$ 
for $\lambda=0.7$
in Fig.~\ref{fit_FM}. It is obvious that the  LSUB$m$ data are well fitted
by
the applied extrapolation function.

Let us briefly discuss the relation of our results to
the schematic phase diagram previously presented in Ref.~\onlinecite{ueda07}.
As stated in Ref.~\onlinecite{ueda07} the approximations used there (bond-operator mean-field theory as well
as a second-order perturbation theory in $\lambda$)
may be not reliable for large $\lambda \sim 1$ and large $J_2 \sim 1$.
The main features of our phase diagram  agree well with those presented in
Ref.~\onlinecite{ueda07}.
For $\lambda \to 0$ we obtain
the same transition point $J_2^{c2}=0.5$ between the QPM and the PFM phases.
However, for $\lambda \sim 1$ and $J_2 \sim 1$ some differences appear.
For instance, we do not find a nematic phase for $\lambda \lesssim 1$, as it was
discussed in
Ref.~\onlinecite{ueda07}.
Note that the absence of the nematic phase
is in agreement with the recent findings for
the uniform model, i.e. at  $\lambda=1$.\cite{richter10,cabra2011}  
Moreover, the second-order transition between the QPM and CAF phases for large
$J_2$ is found in Ref.~\onlinecite{ueda07} at $\lambda \approx 0.5$, but it
should be at $\lambda \approx 0.4$.

\section{Summary}

Inspired by a recent investigation of a frustrated two-dimensional Heisenberg
model proposed to describe the magnetic properties of
(CuCl)LaNb$_2$O$_7$, see Refs.~\onlinecite{kageyama05} and \onlinecite{ueda07}, we investigate   
the GS phase diagram of the spin-$1/2$  $J_1$-$J_2$
Heisenberg model on the square lattice with plaquette structure. 
The 4-site plaquettes  carrying the (strong) intra-plaquette bonds $J_1$ and
$J_2$ are coupled  to each other by (weaker) inter-plaquette bonds $\lambda J_1$ and
$\lambda J_2$, $ 0 \le \lambda \le 1$.
The parameter $\lambda$ can also be thought of modeling a distortion of the
underlying square lattice.   
     
We consider
AFM ($J_1 > 0$) as well as   FM
nearest-neighbor exchange coupling  ($J_1 < 0$). 
Except the phases with magnetic LRO (FM, \Neel, and collinear striped AFM)
and the non-magnetic quantum paramagnetic phase known  from the standard
spin-$1/2$  $J_1$-$J_2$ model we find for FM $J_1$ a new plaquette phase
showing antiferromagnetic long-range of $s=2$ block spins associated with  
4-spin plaquettes.   
For the AFM model ($J_1 > 0$) the region of the quantum paramagnetic
phase is significantly increased for $\lambda < 1$ compared to the
standard model, thus increasing the prospects of finding a magnetically
disordered low-temperature phase in real magnetic
materials, where the exchange pattern may deviate from the standard
$J_1$-$J_2$ model, see, e.g., Refs. \onlinecite{rosner2010,volkova2010} and \onlinecite{janson2010}.

\section*{Acknowledgments}
The research was supported by the DFG (project RI 615/16-2).



\begin{thebibliography}{999}

\bibitem{chandra88} P.~Chandra and B.~Doucot,  { Phys. Rev. B} {\bf 38},
  9335 (1988).

\bibitem{dagotto89} E.~Dagotto and A.~Moreo,  { Phys. Rev. Lett.} {\bf 63},
  2148 (1989).

\bibitem{schulz} H.J.~Schulz and T.A.L.~Ziman, { Europhys. Lett.}
  {\bf 18}, 355 (1992);
H.J. Schulz, T.A.L.~Ziman, and D.~Poilblanc, { J.~Phys.~I}
  {\bf 6}, 675 (1996).
\bibitem{richter93}
   J.~Richter, Phys. Rev. B {\bf 47}, 5794 (1993).
\bibitem{richter94} 
J.~Richter, N.B.~Ivanov, and K.~Retzlaff,
      Europhys. Lett. {\bf 25}, 545 (1994).
\bibitem{zhito96}
M.E. Zhitomirsky and K. Ueda
Phys. Rev. B {\bf 54}, 9007 (1996).


\bibitem{Trumper97}
A.~E.~Trumper, L.~O.~Manuel, C.~J.~Gazza, and H.~A.~Ceccatto,
Phys. Rev. Lett. {\bf 78}, 2216 (1997).
\bibitem{bishop98} R.F.~Bishop, D.J.J.~Farnell, and J.B.~Parkinson,
 {  Phys. Rev. B} {\bf 58}, 6394 (1998).


\bibitem{singh99} 
R. R. P. Singh, Z. Weihong, C. J. Hamer, and J. Oitmaa, Phys. Rev. B {\bf 60}, 7278 (1999).

\bibitem{capriotti01} L.~Capriotti, F.~Becca, A.~Parola, and S.~Sorella, 
{ Phys. Rev. Lett.} {\bf 87}, 097201 (2001).
\bibitem{sushkov01}
O.~P.~Sushkov, J.~Oitmaa, and Z.~Weihong,
Phys. Rev. B {\bf  63}, 104420 (2001).
\bibitem{Sir:2006} J. Sirker, Z. Weihong, O. P. Sushkov, and J. Oitmaa,
Phys. Rev. B {\bf 73}, 184420 (2006).
\bibitem{Schm:2006} D. Schmalfu{\ss}, R. Darradi, J. Richter,
J.~Schulenburg, and D.~Ihle,
Phys. Rev. Lett. {\bf 97}, 157201 (2006).

\bibitem{sousa}
 J.R.~Viana and J.R.~de~Sousa,
 Phys.\ Rev.\ B {\bf 75}, 052403 (2007).

\bibitem{bishop08} R.F.~Bishop, P.H.Y.~Li, R.~Darradi, and J. Richter,
      J. Phys.: Condens. Matter {\bf 20}, 255251 (2008).

\bibitem{xxz} R.F.~Bishop, P.H.Y.~Li, R.~Darradi, J.~Schulenburg and J.
Richter,
     Phys. Rev. B {\bf 78}, 054412 (2008).                    


 \bibitem{darradi08} R.~Darradi, O.~Derzhko, R.~Zinke, J.~Schulenburg,  S.~E.~Kr\"uger, and
       J.~Richter,
       Phys. Rev. B {\bf 78}, 214415 (2008).

\bibitem{ortiz} L.~Isaev, G. Ortiz, and J. Dukelsky,
Phys. Rev. B {\bf 79}, 024409 (2009).

\bibitem{singh2009} T. Pardini and R.R.P. Singh,
Phys. Rev. B {\bf 79}, 094413 (2009).

\bibitem{cirac2009} 
V. Murg, F. Verstraete, and J. I. Cirac, Phys. Rev. B {\bf 79},
195119 (2009).

\bibitem{ED40}J. Richter and J. Schulenburg,
Eur. Phys. J. B {\bf 73}, 117 (2010).

\bibitem{fprg} J. Reuther and P. W\"olfle,  Phys. Rev. B {\bf 81}, 144410 (2010).


\bibitem{cirac2010}
A. Sfondrini, J. Cerrillo, N. Schuch, and J.
I. Cirac, Phys. Rev. B {\bf 81}, 214426 (2010).
\bibitem{balents2011} H.-C. Jiang, H. Yao, L. Balents,
arXiv:1112.2241 (2011).

\bibitem{verstrate2011}  L. Wang, Z.-C. Gu, X.-G. Wen, F. Verstraete,
arXiv:1112.3331 (2011).


\bibitem{kaul04} E.~E.~Kaul, H.~Rosner, N.~Shannon, R.V.~Shpanchenko, and C.~ Geibel, 
J. Magn. Magn. Mater. \textbf{272-276(II)}, 922 (2004).
\bibitem{jmmm07} M. Skoulatos, J.P. Goff, N. Shannon, E.E. Kaul, C. Geibel, A.P. Murani, M.
Enderle, and  A.R. Wildes,
J. Magn. Magn. Mater. \textbf{310}, 1257 (2007).
\bibitem{carretta2009}
P. Carretta,  M. Filibian,  R. Nath,  C. Geibel, and P. J. C. King,
Phys. Rev. B {\bf 79}, 224432 (2009).
\bibitem{enderle}
M. Skoulatos, J.P. Goff, C. Geibel, E.E. Kaul, R. Nath, N. Shannon,
B. Schmidt, A.P. Murani, P.P. Deen, M. Enderle, and A.R. Wildes, Europhys.
Lett. {\bf 88}, 57005 (2009).
\bibitem{nath} R. Nath, Y. Furukawa, F. Borsa, E. E. Kaul, M. Baenitz, C. Geibel, and D. C.
Johnston,
Phys. Rev. B {\bf 80}, 214430 (2009).

\bibitem{rosner09}
A.A. Tsirlin and H. Rosner,
Phys. Rev. B {\bf 79},  214417  (2009).
\bibitem{rosner09a}
 A.A. Tsirlin, B. Schmidt,  Y. Skourski,
   R. Nath, C. Geibel, and H. Rosner,
Phys. Rev. B {\bf 80}, 132407  (2009).
\bibitem{carretta2011} L. Bossoni, P. Carretta, R. Nath, M. Moscardini,
M. Baenitz, and C. Geibel, Phys. Rev. B {\bf 83}, 014412 (2011).

\bibitem{nath2008}
R. Nath, A.A. Tsirlin, H. Rosner, and C. Geibel,
Phys. Rev. B {\bf 78}  064422 (2008).



\bibitem{shannon04}
N. Shannon, B. Schmidt, K. Penc, and P. Thalmeier, Eur. Phys. J.  B {\bf 38}, 599 (2004).

\bibitem{shannon06}
N. Shannon, T. Momoi, and P. Sindzingre, Phys. Rev. Lett. {\bf 96}, 027213 (2006).

\bibitem{sindz07}
P. Sindzingre, N. Shannon and T. Momoi, J. Magn. Magn. Mat.
{\bf  310}, 1340 (2007).
\bibitem{schmidt07} B.~Schmidt, N.~Shannon, and P.~Thalmeier, J. Phys.:
Condens. Matter \textbf{19}, 145211 (2007).
\bibitem{schmidt07_2} B.~Schmidt, N.~Shannon, and P.~Thalmeier, J. Magn. Magn. Mater. \textbf{310}, 1231
(2007).
\bibitem{shannon09} P. Sindzingre, L. Seabra, N. Shannon, and
T. Momoi,
J. Phys.: Conf.  Series {\bf 145}, 012048 (2009).
\bibitem{sindz09}  P. Sindzingre, N. Shannon, and
T. Momoi, J. Phys.: Conf.  Series {\bf 200}, 022058 (2010).
\bibitem{haertel10}
M. H\"artel, J. Richter, D.~Ihle, and S.-L.~Drechsler,
	Phys. Rev. B {\bf 81}, 174421 (2010).

\bibitem{richter10}
J. Richter, R. Darradi, J. Schulenburg, D. J. J. Farnell, and H. Rosner, Phys. Rev. B {\bf 81}, 174429 (2010).

\bibitem{momoi2011}
R. Shindou, S. Yunoki, and T. Momoi,
Phys. Rev. B {\bf 84}, 134414 (2011).
\bibitem{cabra2011}
H. Feldner, D. C. Cabra, and G. L. Rossini,
Phys. Rev. B {\bf 84}, 214406 (2011).


\bibitem{hjs}
H.-J. Schmidt, J. Phys. A \textbf{38}, 2123 (2005).




\bibitem{voigt02} 
A. Voigt, Comp. Phys. Comm. {\bf 146}, 125 (2002).

\bibitem{ueda07} 
H.T. Ueda and K. Totsuka, Phys. Rev. {\bf B 76}, 214428 (2007).

\bibitem{fabricio08}
A. F. Albuquerque, M. Troyer, and J. Oitmaa, Phys. Rev. B {\bf 78}, 132402 (2008).

\bibitem{wenzel08} 
S. Wenzel, L. Bogacz, and W. Janke, Phys. Rev. Lett. {\bf 101}, 127202 (2008).

\bibitem{wenzel09}
S. Wenzel and W. Janke, Phys. Rev. B {\bf 79}, 014410 (2009).

\bibitem{fledderjohann09} 
A. Fledderjohann, A. Kl\"umper, and K.-H. M\"utter, 
Eur. Phys. J. B. {\bf 72}, 551 (2009); Eur. Phys. J. B. {\bf 72}, 559 (2009).

\bibitem{rosner2010}
A.A. Tsirlin, R. Nath, A.M. Abakumov, R.V.
Shpanchenko, C. Geibel, and H. Rosner, 
Phys. Rev. B {\bf 81}, 174424 (2010).

\bibitem{kageyama05} 
H. Kageyama, T. Kitano, N. Oba, M. Nishi, S. Nagai, K. Hirota, L. Viciu, J. B. Wiley, J. Yasuda, 
Y. Baba, Y. Ajiro, and K. Yoshimura, J. Phys. Soc. Jpn. {\bf 74}, 1702 (2005).


\bibitem{rosner_co}
A. A. Tsirlin, A. M. Abakumov, G. Van Tendeloo, and H.
Rosner, Phys. Rev. B {\bf 82}, 054107  (2010),
A.A. Tsirlin and H. Rosner,
Phys. Rev. B {\bf 82},  060409 (2010).

\bibitem{spinpack}
{\tt http://www-e.uni-magdeburg.de/jschulen/spin/}.

\bibitem{wir04}  J. Richter, J. Schulenburg and A. Honecker,
in {\em Quantum Magnetism},
eds U.~Schollw{\"{o}}ck, J.~Richter, D.J.J.~Farnell, and R.F.~Bishop,
Lecture Notes in Physics {\bf 645} (Springer-Verlag,
Berlin, 2004), p. 85.

\bibitem{bounce} D.J.J.~Farnell, R.~Darradi, R.~Schmidt, and J.~Richter,
      Phys. Rev. B {\bf 84}, 104406 (2011).


\bibitem{li2012}   P.H.Y. Li, R.F. Bishop, D.J.J. Farnell, J. Richter, and C. E.
       Campbell,
       Phys. Rev. B {\bf 85}, 085115 (2012).

\bibitem{krueger00} 
S.~E.~Kr\"uger,  J.~Richter, J.~Schulenburg, D.~J.~J.~Farnell, and R.~F.~Bishop,
Phys. Rev. B {\bf 61}, 14607 (2000).

\bibitem{krueger01} 
S.~E.~Kr\"uger and  J.~Richter,
Phys. Rev. B {\bf 64}, 024433 (2001).



\bibitem{farnell05} D.J.J.~Farnell,  J.~Schulenburg, J.~Richter, and
      K.A.~Gernoth,
      Phys. Rev. B {\bf 72}, 172408 (2005)


\bibitem{farnell09} D.J.J. Farnell, J. Richter; R. Zinke; R.F. Bishop,
      J. Stat. Phys., {\bf 135}, 175 (2009)



\bibitem{zeng98}
C.~Zeng, D.~J.~J.~Farnell, and R.~F.~Bishop,
J. Stat. Phys. {\bf 90}, 327 (1998).

\bibitem{bishop98a} R. F. Bishop, in {\it Microscopic Quantum Many-Body Theories and Their
Applications}, edited by
   J. Navarro and A. Polls, Lecture Notes in Physics 510 (Springer-Verlag,
Berlin, 1998), p.1.

\bibitem{bishop00} 
R.~F.~Bishop, D.~J.~J.~Farnell, S.~E.~Kr\"uger,  J.~B.~Parkinson, J.~Richter,
and  C. Zeng,
J. Phys.: Condens. Matter { \bf 12}, 6887 (2000).

\bibitem{bishop04}
D.~J.~J.~Farnell and R.~F.~Bishop,
in {\it Quantum Magnetism}, Lecture Notes in Physics Vol. {\bf 645}, 
edited by U.~Schollw\"ock, J.~Richter, D.~J.~J.~Farnell, and R.~F.~Bishop (Springer, Berlin, 2004), p. 307. 


\bibitem{singh88} 
R.R.P. Singh and M.P. Gelfand, Phys. Rev. Lett. {\bf 61}, 2484 (1988).

\bibitem{ivanov96} N.B. Ivanov,  S.E. Kr\"uger, and J. Richter, 
      Phys. Rev. B {\bf 53}, 2633 (1996).



\bibitem{vidal09}
B. Bauer, G. Vidal, and  M. Troyer,
J. Stat. Mech. P09006 (2009).

\bibitem{volkova2010}
O. Volkova, I. Morozov, V. Shutov, E. Lapsheva, P. Sindzingre, O. Cepas, M. Yehia, V. Kataev, 
R. Klingeler, B. B\"uchner, and A. Vasiliev,
Phys. Rev. B {\bf 82}, 054413 (2010).

\bibitem{janson2010} 
O. Janson, A. A. Tsirlin, and H. Rosner,
Phys. Rev. B {\bf 82}, 184410 (2010).





\end{thebibliography}
\end{document}